\documentclass[]{raa_twocolumn}           
\usepackage{graphicx,times}
\usepackage{natbib}
\usepackage{amssymb,amsmath}
\bibpunct{(}{)}{;}{a}{}{,}

\def\mc{{\mathcal{M}}_{\rm c}}
\def\gpcyr{\rm \,Gpc^{-3}\,yr^{-1}}

\graphicspath{{figs/}}

\usepackage[a4paper=true,pagebackref=true]{hyperref}
\hypersetup{pdftitle = The title of my PDF, pdfauthor = My name, pdfsubject= The subject, pdfkeywords = keyword1 keyword2 keyword3} 
\hypersetup{colorlinks = true, linkcolor = green, anchorcolor = red, citecolor = blue, filecolor = red, pagecolor = red, urlcolor = red}

\begin{document}

   \title{On Detecting Stellar Binary Black Holes via the LISA-Taiji Network}

 \volnopage{ {\bf 20XX} Vol.\ {\bf X} No. {\bf XX}, 000--000}
   \setcounter{page}{1}

   \author{Ju Chen \inst{1,2}, Changshuo Yan $^\ast$\inst{1,2}, Youjun Lu \inst{1,2}, Yuetong Zhao \inst{1,2}, Junqiang Ge \inst{1}  }

   \institute{ National Astronomical Observatories, Chinese Academy of Sciences, 20A Datun Road, Beijing 100101, China; {\it $^\ast$\,yancs@nao.cas.cn}\\
        \and
             University of Chinese Academy of Sciences, 19A Yuquan Road, Beijing 100049, China\\
\vs \no
   {\small Received 20XX Month Day; accepted 20XX Month Day}
}

\abstract{The detection of gravitational waves (GWs) by ground-based laser interferometer GW observatories (LIGO/Virgo) reveals a population of stellar binary black holes (sBBHs) with (total) masses up to $\sim 150M_\odot$, which are potential sources for space-based GW detectors, such as LISA and Taiji. In this paper, we investigate in details on the possibility of detecting sBBHs by the LISA-Taiji network in future. We adopt the sBBH merger rate density constrained by LIGO/VIRGO observations to randomly generate mock sBBHs samples. Assuming an observation period of $4$ years, we find that the LISA-Taiji network may detect several tens (or at least several) sBBHs with signal-to-noise ratio (SNR) $>8$ (or $>15$), a factor $2-3$ times larger than that by only using LISA or Taiji observations. Among these sBBHs, no more than a few that can merge during the $4$-year observation period. If extending the observation period to $10$ years, then the LISA-Taiji network may detect about one hundred (or twenty) sBBHs with SNR $>8$ (or $>15$), among them about twenty (or at least several) can merge within the observation period. Our results suggest that the LISA-Taiji network may be able to detect at least a handful to twenty or more sBBHs even if assuming a conservative SNR threshold ($15$) for ``detection'', which enables multi-band GW observations by space and ground-based GW detectors. We also further estimate the uncertainties in the parameter estimations of the sBBH systems ``detected'' by the LISA-Taiji network. We find that the relative errors in the luminosity distance measurements and sky localization are mostly in the range of $0.05-0.2$ and $1-100\deg^2$, respectively, for these sBBHs.
}

   \authorrunning{J. Chen et al. }            
   \titlerunning{LISA-Taiji Network}  
   \maketitle

%
\section{Introduction}
\label{sec:intro}

Detection of gravitational waves (GWs) by ground-based laser interferometer GW observatories reveals a population of massive stellar binary black holes (sBBHs) with primary component masses from $30M_\odot$ up to $80 M_\odot$ \citep{Abbott2016c, Abbott2019, Abbott2020, ligoO3a}. These massive stellar black holes (BHs) were not found before by electromagnetic (EM) observations (e.g., from X-ray binary observations)  \citep[e.g.][]{Ozel2010, Corral-Santana2016, Casares2017} except for one 68 $M_\odot$ black hole in binary system claimed by \citet{Liu2019} recently, and are hard to be formed directly via stellar evolution \citep[e.g.,][]{Belczynski2002, Woosley2002}. Low-metallicity environments \citep[e.g.,][]{Abbott2016e, Belczynski2016} or dynamical formation mechanisms \citep[e.g.,][]{Rodriguez2016, Mapelli2016} are considered as possible explanation.

Some of the massive sBBHs may radiate GWs at the low frequency band ($\sim 10^{-3}-1$\,Hz) and evolve to a final coalescence within several years. Therefore, they are potential sources for multiband GW observations \citep{Sesana2016, Liu2020}, i.e., by both future space GW detectors (the Laser Interferometer Space Antenna (LISA) \citep{Amaro-Seoane2017}, Taiji \citep{Ruan2018}, and Tianqin \citep{Liu2020c}) at the low frequency band and the ground-based GW detectors, such as LIGO/VIRGO/KAGRA \citep{LIGO2015, Virgo2015, Kagra2019}, Einstein Telescope (ET, \citet{ET2010}), and Cosmic Explorer (CE, \citet{CE2017}), at high frequency band. The number of such sources was predicted to be tens to hundreds \citep{Sesana2016, Sesana2017, Gerosa2019, Liu2020}, which suggests multiband GW observation is quite promising. However, \citet{Moore2019} recently cautioned that such multiband GW observations is challenging as the signal-to-noise ratio (SNR) threshold to claim a ``detection'' at the LISA band may substantially higher than it was expected, especially for those ``merging'' sBBHs, due to the complexity of signal space, and the expected number of sBBHs for multiband GW observations may be limited or even negligible within the mission time of LISA. This raised a significant problem for the possibility of multi-band GW observations of sBBHs and its mergers.  

LISA and Taiji, both a trio consisting of three spacecraft with arm length of $2.5$/$3$ millions kilometers, are planning to launch in $2030$-$2035$ and expected to have an overlap of observation periods in its mission schedule \citep{Ruan2020a}. LISA and Taiji are both on heliocentric orbits, with the former one behind the Earth by $20^\circ$ and the latter one ahead of the Earth by $20^\circ$, and they may form a network for low frequency GW observations. With this network, not only the SNR of a GW event detected by both instruments, but also the sky localization of the event can be improved a lot. This has been demonstrated by detailed analysis in \citet{Ruan2021} and \citet{Wang2020a} for massive binary black holes, and the potential application in cosmology was discussed in \citet{Wang2020}. It is of great interest to ask whether such a network can also improve the capability of space GW detectors to catch the GW signals from sBBHs and enable the multi-band GW observations of at least a handful sBBHs. In this paper, we analyze the potential of the LISA-Taiji network to detect inspiralling and merging sBBHs and estimate the number of sBBHs that can be detected by it. We also further estimate the uncertainties of the parameters measured from GW signals of those sBBH systems observed by the LISA-Taiji network. 

The paper is organized as follows. In Section~\ref{sec:mockdata}, we describe the processes to generate mock samples of sBBHs at low redshift for LISA and Taiji to observe. In Sections~\ref{sec:SNR} and \ref{sec:uncertainties}, we introduce the methods to estimate SNR and uncertainties in the parameter extractions from LISA, Taiji, and LISA-Taiji network ``observations'' of the mock sBBHs by using Fisher information matrix method. We present our results in Section~\ref{sec:results}. Conclusions are summarized in Section~\ref{sec:conclusions}.

Throughout the paper, we adopt the Planck cosmology with $(h_0, \Omega_{\rm m}, \Omega_{\Lambda}) = (0.673, 0.315, 0.685)$ and $h_0=H_0/100{\rm km\,s^{-1}\,Mpc^{-3}}$ \citep{PlanckCollaboration2014}. Here $H_0$ is the Hubble constant, $\Omega_{\rm m}$ and $\Omega_{\Lambda}$ are the fractions of present cosmic density contributed by matter and cosmological constant, respectively. 

\section{Mock catalog for sBBHs}
\label{sec:mockdata}

The merger rate density of sBBHs may be described by a power-law at low redshift as $R(z) = R_0(1+z)^\kappa$, where $R_0$ is the local merger rate density. LIGO/VIRGO observations have put a strong constraint on $R_0$, and it is $19.1^{+16}_{-10}\gpcyr$  (or $23.9^{+14.9}_{-8.6}\gpcyr$ if without consideration of the redshift evolution) \citep{ligoO3a}. According to LIGO/VIRGO observations, the probability distributions of primary mass $m_1$ and mass ratio $q$ are also constrained,  with $m_1$ and $m_2=q m_1$ denoting the masses of the two components. Here we adopt the power-law plus a peak model  $P(m_1, q)$ in \citet{ligoO3a} for primary mass distribution and mass ratio distribution (see Eqs. B5-B8 in the paper).

With the above settings, the distributions of sBBHs with various parameters can be given by \citep[e.g., see also][]{ZhaoLu2021}
\begin{equation}
\frac{d^4N}{df dz dm_1 dq} = R(z)P(m_1,q)\frac{dt}{df}\frac{1}{1+z}\frac{dV}{dz} ,
\label{eq:num-circ}
\end{equation}
where $dt/df=(5/96)\pi^{-8/3}(G\mc/c^3)^{-5/3}$ with $\mc= (1+z) M_{\rm c}$ representing the redshifted chirp mass and $M_{\rm c} = m_1q^{3/5}(1+q)^{-1/5}$ representing the chirp mass. Since the lifetimes of sBBHs with frequency $f$ considered in this paper are small (typically less than a few hundreds years), we ignore the slight redshift difference between the inspiralling sBBHs with frequency $f$ and its later merger time. 

\section{SNR estimation for sBBHs}
\label{sec:SNR}

For each individual GW source, the SNR ($\varrho$) for the LISA-Taiji network may be estimated by
\begin{eqnarray}
\label{eq:snr}
\varrho^2 = \sum_{j=1}^{4}  \int_{f_{\rm i}}^{f_{\rm f}} \frac{4\tilde{h}_j^{*}(f) \tilde{h}_j(f)}{S_{n,j}(f)} d f,
\end{eqnarray}
where $j$ refers to independent detectors, ``$1$" and ``$2$" refer to the two Michelson interferometers of LISA ($S_{\rm n,1}=S_{\rm n,2}=S_{\rm n,LISA}$) while ``$3$" and ``$4$" refers to the Taiji ($S_{\rm n,3}=S_{\rm n,4}=S_{\rm n,Taiji}$). $S_{\rm n}(f)$ is the non-sky-averaged single-detector noise power density, we take the sensitivity curve for LISA from \citet{Robson2019} and that for Taiji from \citet{Ruan2020}, then convert them into non-sky average noise spectrum \citep{Liu2020}. Adopting the Newtonian approximation for the inspiral stage of BBHs \citep{Maggiore2008}, the Fourier transform of GW signal $h(t)$ can be written as
\begin{equation}
\tilde{h}(f)=\left(\frac{5}{24}\right)^{1 / 2} \frac{1}{\pi^{2 / 3}} \frac{c}{d_{\rm L}}\left(\frac{G \mc}{c^{3}}\right)^{5 / 6} f^{-7 / 6} Q e^{i \Psi(f)}.
\label{eq:h_f}
\end{equation}
Here $c$ is the speed of light, $G$ the gravitational constant, $d_{\rm L}$ the luminosity distance, $\Psi(f)$ the strain phase, $Q$ a quantity related to the detector's pattern function $F^+$ and $F^\times$, i.e.,
\begin{eqnarray}
Q(\theta, \phi, \psi ; \iota)& =& \sqrt{F^2_{+} \left(\frac{1+\cos ^{2} \iota}{2}\right)^2+ F^2_{\times} \cos^2 \iota}, 
\end{eqnarray}
with the detector's pattern function
\begin{eqnarray}
F_{+}(\theta, \phi, \psi)& =& \frac{1}{2}\left(1+\cos ^{2} \theta\right) \cos 2 \phi \cos 2 \psi \nonumber \\
& &-\cos \theta \sin 2 \phi \sin 2 \psi, \\
F_{\times}(\theta, \phi, \psi) &=& \frac{1}{2}\left(1+\cos ^{2} \theta\right) \cos 2 \phi \sin 2 \psi \nonumber \\
&& +\cos \theta \sin 2 \phi \cos 2 \psi.
\end{eqnarray}
Note here $\theta$, $\phi$, $\psi$, and $\iota$ are the polar angle, azimuthal angle, polarization angle, and the inclination angle between the sBBH angular momentum and the vector pointing from detector to source in the detector's frame, respectively, in the detector's frame. In considering the limited numbers of detection for each realization, we would use average SNR in our study, which should be more representative.

Averaging over all possible directions and inclinations, we have
\begin{equation}
\left\langle|Q(\theta, \phi, \psi ; \iota)|^{2}\right\rangle^{1 / 2}=\frac{2}{5}.
\end{equation}

Combining equations~\eqref{eq:snr} and \eqref{eq:h_f}, the averaged SNR may be given by
\begin{eqnarray}
\varrho & = &
\frac{\sqrt{2}}{\sqrt{15} \pi^{\frac{2}{3}}}
\frac{c}{d_{\rm L}}\left(\frac{G\mc}{c^{3}}\right)^{\frac{5}{6}} 
\left[\sum_{j=1}^{4}\int_{f_{\rm i}}^{f_{\rm f}} df \frac{f^{-\frac{7}{3}}}{S_{n,j}(f)}\right]^{\frac{1}{2}}, \nonumber \\
\label{eq:rho}
\end{eqnarray}
where $f_{\rm i}$, assigned according to the frequency distribution in equation~\eqref{eq:num-circ}, is the initial GW frequency in the observer's frame of the mock sBBH at the beginning of LISA/Taiji observation, $f_{\rm f}= \min(f_{\rm end}, f_{\rm ISCO}, 1{\rm Hz})$, where $1$Hz is the upper frequency limit of the observations and $f_{\rm ISCO}= 2.2M_\odot/(M(1+z))$\,kHz, $M=m_1+m_2$ is the total mass of BBH. While  $f_{\rm end}=(1/8\pi)(G\mc/c^3)^{5/8}((T_{\rm i}-T_{\rm obs})/5.0)^{-3/8}$ denotes the GW frequency at the end of observation if assuming the duration of the mission is $T_{\rm obs}=4$ years, where $T_{\rm i}=5(8\pi f_{\rm i})^{-8/3}(G\mc/c^3)^{-5/3}$.

We take the non-sky-averaged waveform in the parameter estimation. The sBBH orbital motions during the observation period are functions of time $t$. We change the detector frame parameters $(\theta,\phi,\psi)$ to the ecliptic coordinates.  With the source position denoted by $(\theta_{\rm S},\phi_{\rm S})$ and the orbit angular momentum direction of the mock BBH denoted by $(\theta_{\rm L},\phi_{\rm L})$, we can then use $(\theta_{\rm S},\phi_{\rm S},\theta_{\rm L},\phi_{\rm L})$ to replace $(\theta,\phi,\psi,\iota)$ (see Eqs. (10-19) in \citet{Liu2020}  or Eqs. (3.16-3.22) in \citet{Cutler1998} ). In the frame transformation, there are two important parameters: (1) the azimuthal angle of the detector around the Sun $\Phi(t)=\Phi_0+\frac{2\pi t(f)}{T}$, $T$ is the orbital period of the detector, equals to one year; and (2) the initial orientation of the detector arms $\alpha_0$.

When estimating the SNR for LISA or Taiji observations only, we also adopt equation~\eqref{eq:rho} but set $j$ from $1$ to $2$ and from $3$ to $4$ in the summation for LISA and Taiji, respectively.

\section{Uncertainties in parameter Estimations of sBBH systems}
\label{sec:uncertainties}

The expected uncertainties in the measurements of the BBH parameters $\boldsymbol{\Xi} =\left\{d_{\rm L}, \mc, \eta, t_{\rm c}, \phi_{\mathrm{c}}, \theta_{\rm S}, \phi_{\rm S}\right\}$ may be estimated by using the Fisher matrix method. The Fisher matrix can be obtained as
\begin{eqnarray}
\Gamma_{a b} & = & \left(\left.\frac{\partial h}{\partial \boldsymbol{\Xi}^{a}} \right| \frac{\partial h}{\partial \boldsymbol{\Xi}^{b}}\right) \nonumber \\
& = & \sum_{j=1}^{4} 2\int_{f_{\rm i}}^{f_{\rm f}} \frac{\frac{\partial \tilde{h}_j^{*}(f)}{\partial \boldsymbol{\Xi}^{a}}\frac{\partial \tilde{h}_j(f)}{\partial \boldsymbol{\Xi}^{b}}+\frac{\partial \tilde{h}_j^{*}(f)}{\partial \boldsymbol{\Xi}^{b}}\frac{\partial \tilde{h}_j(f)}{\partial \boldsymbol{\Xi}^{a}}}{S_{n,j}(f)} df,
\end{eqnarray}
where $j$ refers to detector ``1" (with $\alpha_0=0$, $\Phi_0=0$) or ``2" (with $\alpha_0=\pi/4$,  $\Phi_0=0$) or ``3" (with $\alpha_0=\Delta\alpha$, $\Phi_0=2\pi/9$) or ``4" (with $\alpha_0=\Delta\alpha+\pi/4$,  $\Phi_0=2\pi/9$). Here $\Delta\alpha$ is the difference of the initial orientation between LISA and Taiji and this value will not affect the result much, so we set $\Delta\alpha= \pi/2$. Because LISA is set to be in a heliocentric orbit behind the Earth by about $20^{\circ}$ while Taiji in a heliocentric orbit ahead of the Earth by about $20^{\circ}$, so we set $\Phi_0=0^\circ$ for LISA and $\Phi_0=2\pi/9$ for Taiji.

Given $\Gamma^{ab}$, then we have
\begin{equation}
\left\langle\delta \Xi^{a} \delta \Xi^{b}\right\rangle=\left(\Gamma^{-1}\right)^{a b},
\end{equation}
and thus we can estimate the uncertainties in the measurements of $\boldsymbol{\Xi}$ as
\begin{equation}
\Delta \boldsymbol{\Xi}^{a}=\sqrt{\left(\Gamma^{-1}\right)^{a a}}.
\end{equation}
Specifically, the angular resolution $\Delta\Omega$ is defined as
\begin{equation}
\Delta\Omega=2\pi|\sin\theta_{\rm S}|\sqrt{(\Delta\theta_{\rm S}\Delta\phi_{\rm S})^2-\left\langle\Delta\theta_{\rm S}\Delta\phi_{\rm S}\right\rangle^2}.
\end{equation}

The GW strain signal is described by equation~\eqref{eq:h_f} for both detectors. The GW strain phase evolution includes the polarization modulation ($\phi_{\rm p}$) and Doppler modulation ($\phi_{\rm D}$), i.e.,
\begin{eqnarray}
%
%
\Psi(f)&=& 2 \pi f t_{\rm c}-\phi_{\rm c}-\pi/4-\phi_{\rm p}-\phi_{\rm D}+\frac{3}{4}(8 \pi \mc f)^{-5/3} \nonumber \\
& & \times\left[1+\frac{20}{9}\left(\frac{743}{336}+\frac{11\eta}{4}\right) x-16\pi x^{3/2}\right],
%
%
\end{eqnarray}

where
$x(f) \equiv[\pi M(1+z) f]^{2 / 3}$, $\eta=m_1m_2/M^2$ is the symmetric mass ratio, $t_{\rm c}$ and $\phi_{\rm c}$ are the time and the orbital phase at coalescence, respectively.
The polarization modulation can be written as
\begin{equation}
\phi_{\rm p}(t(f))=\arctan \frac{-2 \cos \imath F_{\times}(t(f))}{\left(1+\cos ^{2} \iota\right) F_{+}(t(f))}.
\end{equation}
The motion of the detectors around helio-center causes Doppler modulation of the GW phase as
\begin{equation}
\phi_{\rm D}(t(f))=2 \pi f R \sin \theta_{\rm S} \cos \left(\Phi(t(f))-\phi_{\rm S}\right),
\end{equation}
where $R=1$\,AU is the distance from the detector to the Sun and
\begin{equation}
\begin{aligned}
t(f)=& t_{\rm c}-5(8 \pi f)^{-8 / 3}\mc^{-5 / 3} \\
& \times\left[1+\frac{4}{3}\left(\frac{743}{336}+\frac{11 \eta}{4}\right) x-\frac{32 \pi}{5} x^{3 / 2}\right].
\end{aligned}
\end{equation}

\section{Results}
\label{sec:results}

According to the procedures listed in Section~\ref{sec:mockdata}, we generate $100$ realizations of sBBHs at redshift $z<0.4$. For each of the sBBHs in these realizations, the redshift $z$, primary mass $m_1$, mass ratio $q$, and GW frequency at the starting time of observation $f_{\rm i}$ are assigned according to Equation \ref{eq:num-circ}, and the angles  $(\theta_{\rm S},\phi_{\rm S},\theta_{\rm L},\phi_{\rm L})$ are set to be uniform distribution in the corresponding range.  We then estimate the expected SNR for such an sBBH system observed by LISA, Taiji, or the LISA-Taiji network, for any given observation period. With a given SNR threshold ($\varrho_{\rm th}$) to define the ``detection'', we can get the total number of sBBHs observed by LISA, Taiji, or LISA-Taiji network with SNR above this threshold, i.e., $N(\varrho>\varrho_{\rm th})$. We then rank the $100$ realizations by $N(\varrho>\varrho_{\rm th})$ from low to high and adopt the one in the order of $50$-th, $16$-th, or $84$-th to get the median value and $68\%$ confidence interval of the expected number of ``detectable'' sBBHs. We also estimate the uncertainties in the measurements of system parameters for these sBBHs and the sky localization using the Fisher matrix method introduced above in Section~\ref{sec:uncertainties}. For simplicity, we consider non-spinning sBBHs on circular orbits. Below, we summarize the main results obtained from our calculations. 

\begin{figure}
\centering
\includegraphics[width=0.5\textwidth]{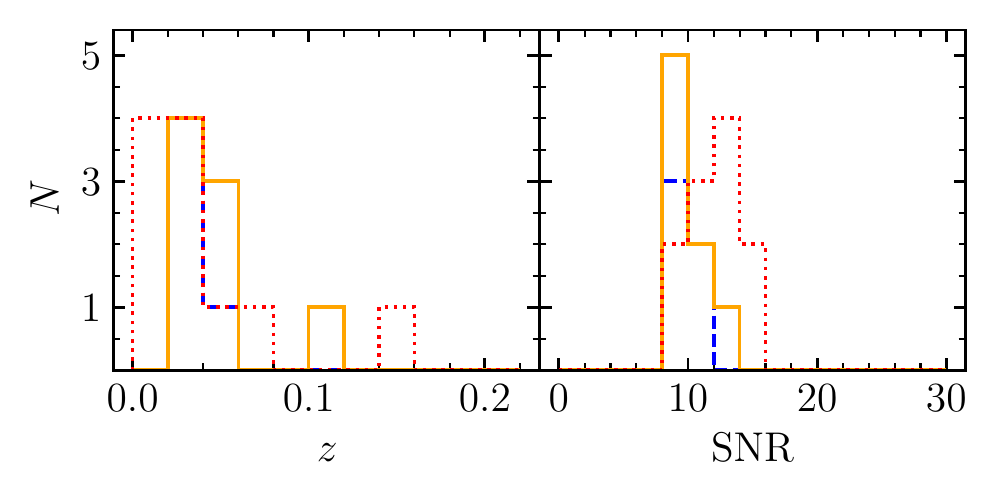}
\caption{Redshift (left panel) and SNR (right panel) distributions of mock sBBH systems that may be ``detected'' with SNR$>8$ by LISA within a $4$-year observation period. Blue dashed, orange solid,  and red dotted lines show the distributions obtained from three different randomly generated samples. These three samples are selected from $100$ realizations of sBBHs at $z<0.4$, ranked by the total number (from low to high) of ``detectable'' sBBHs with SNR$>8$, with rank of $16$-th, $50$-th, and $84$-th, respectively. For those different realizations with the same total number in the ranking, their orders are randomly assigned.
}
\label{fig:SNR_L}
\end{figure}

\begin{figure}
\centering
\includegraphics[width=0.5\textwidth]{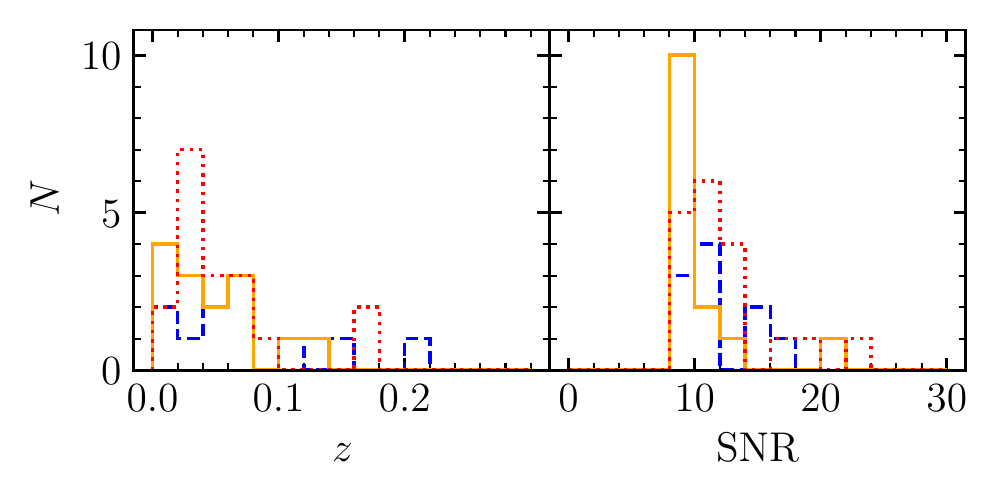}
\caption{Legend similar to Fig.~\ref{fig:SNR_L} but estimated for Taiji with $4$-year observations.}
\label{fig:SNR_T}
\end{figure}

\begin{figure}
\centering
\includegraphics[width=0.5\textwidth]{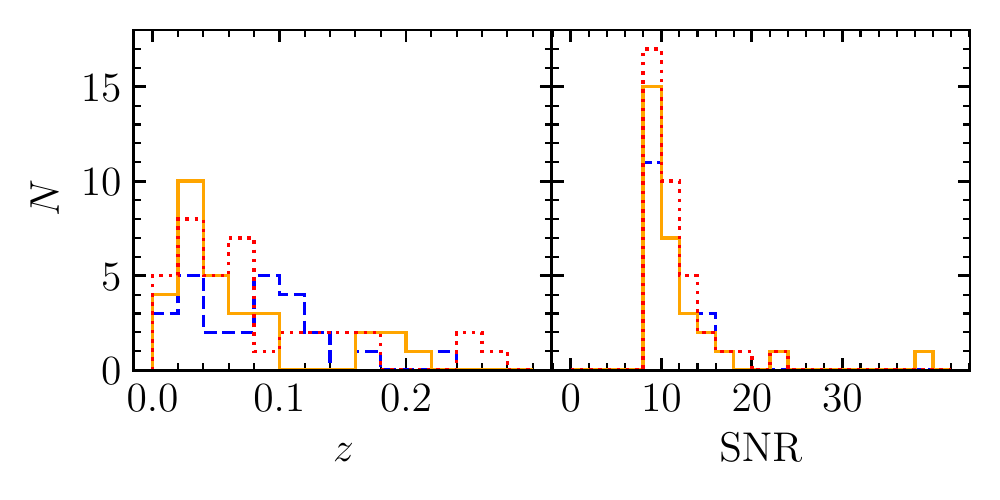}
\caption{
Legend similar to Fig.~\ref{fig:SNR_L} but estimated for the LISA-Taiji network with $4$-year observations.
}
\label{fig:SNR_LT}
\end{figure}

\begin{figure}
\centering
\includegraphics[width=0.5\textwidth]{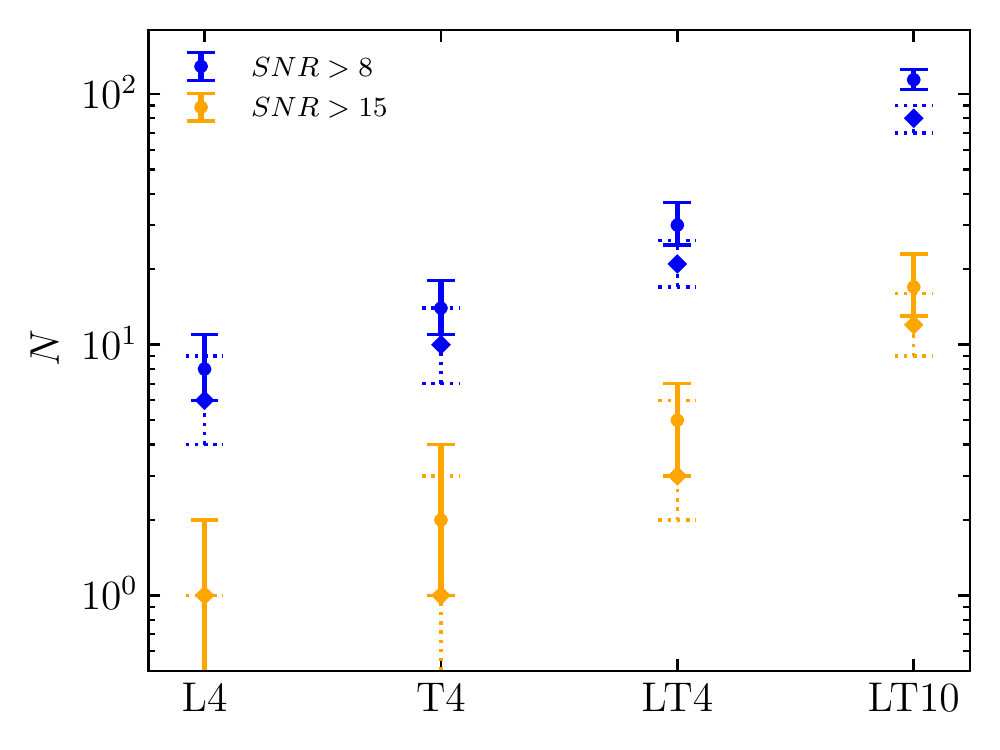}
\caption{
The expected total number of sBBHs that can be ``detected'' with sky-averaged SNR $>8$ and $>15$ by LISA, Taiji, and the LISA-Taiji network. L4, T4, and LT4 denote observations by LISA, Taiji, and the LISA-Taiji network with a $4$-year observation period, and LT10 denotes observations by the LISA-Taiji network over a $10$-year period. The filled circles and its associated error-bars represent the median value and $68\%$ confidence interval of the numbers obtained from $100$ realizations. 
For comparison, the results obtained by using non-averaged SNR are indicated by diamond symbols with dotted line error bars.}
\label{fig:num}
\end{figure}

Figure~\ref{fig:SNR_L} shows the redshift (left panel) and SNR (right panel) distributions of those sBBHs that observed by LISA with SNR threshold $\varrho_{\rm th}=8$ over a $4$-year observation period obtained from three different realizations, with the rank of $16$-th, $50$-th, and $84$-th, respectively. Figures~\ref{fig:SNR_T} and \ref{fig:SNR_LT} also similarly show these distributions for sBBHs observed by Taiji and the LISA-Taiji network, respectively. The median value and $68\%$ confidence interval of the event numbers $N(\varrho>8)$ for different detectors are shown in figure~\ref{fig:num}. As seen from this figure, with LISA alone, the expected number of detection is $8_{-2}^{+3}$, while Taiji can detect slightly more ($14_{-3}^{+4}$).\footnote{The numbers obtained here for LISA/Taiji are smaller than those obtained in \citet{Sesana2016} and \citet{ZhaoLu2021}, which is partly caused by the adoption of a smaller local merger rate density and partly caused by the adoption of a different chirp distribution (or primary mass distribution), based on the latest constraints given by the LIGO/VIRGO O3a observations.} By combining LISA and Taiji together as a network, then the number is more than doubled, up to $30_{-5}^{+7}$. 

Note here that we set the detection SNR threshold for the LISA-Taiji network as $8$, the same as that for a single detector, which may be considered as an optimistic choice. We may also set a conservative SNR threshold as $12$ (or $8\sqrt{2}$) for detector network, similar to earth-based detector network, which roughly means the GW event being detected by both detectors with SNR threshold of $8$. In this case, the combination of two detectors would have no improvement on the detection rate.

If we set the SNR threshold for ``detection'' as $\varrho_{\rm th}=15$, then the expected numbers of sBBH ``detection'' decrease significantly (by a factor of $6\sim8$) as shown in figure \ref{fig:num}. Among the $100$ realizations, only $8$/$34$ realizations can have more than $2$ sBBH ``detections'' for LISA/Taiji, while most realizations ($83$ out of $100$) can have more than $2$ sBBH ``detections'' for the LISA-Taiji network. It is clear that it is not optimistic to eventually detect any sBBHs by only LISA (consistent with \citet{Moore2019}) or only Taiji, though Taiji seems to give a few ``detections'' of sBBHs. If combining LISA and Taiji as a network, then the expected number becomes $5^{+2}_{-2}$. This suggests that the LISA-Taiji network can enable at least a handful ``detections'' of sBBHs.

Note here we adopt the sky-averaged SNR by considering the limited number of detection for each realization. One may also consider the non-averaged SNR and obtain the number of detectable events, as indicated by the diamond symbols with dotted error bar in Figure~\ref{fig:num}. In this case, the predicted detection number of GW events is slightly lower than that from the case by using the sky-averaged SNR.

\begin{figure}
\centering
\includegraphics[width=0.5\textwidth]{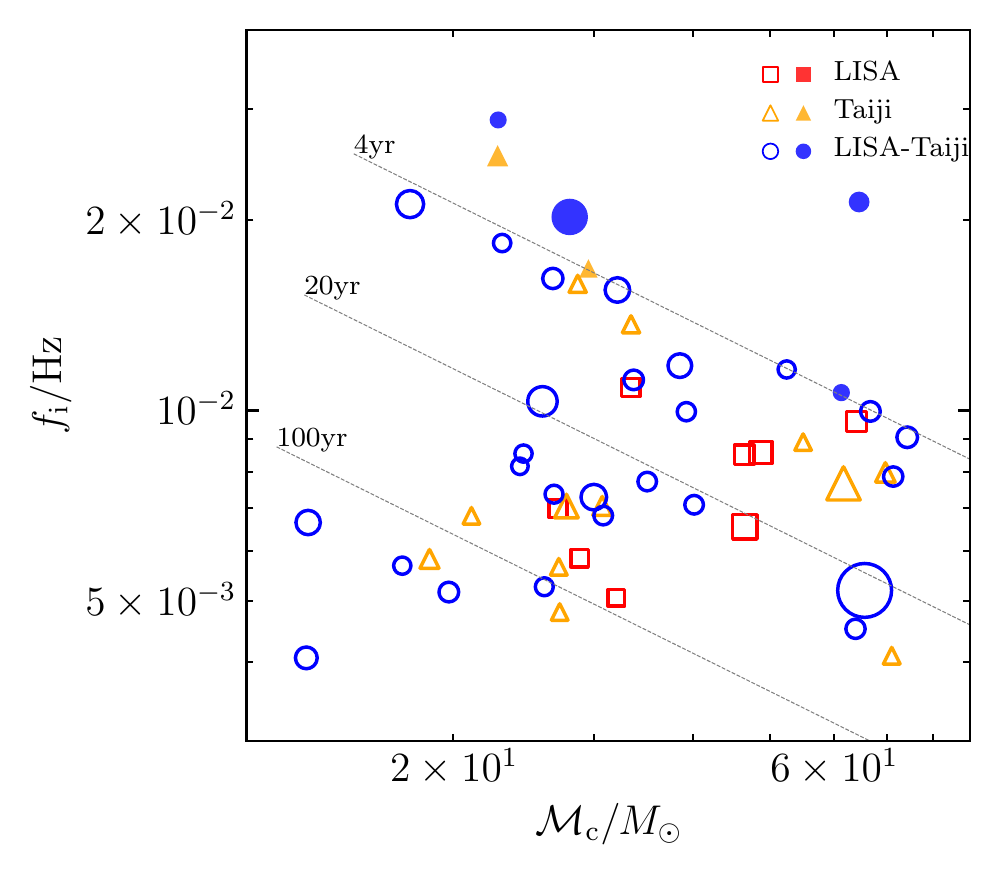}
\caption{
Distribution of mock sBBHs, ``detected'' by LISA-Taiji network with SNR$>8$ on the plane of (redshifted) chirp mass $\mathcal{M}_{\rm c}$ versus observed GW frequency at the starting time of the observation.  Red squares, orange triangles, and blue circles represent those sources ``detected'' by LISA, Taiji, and LISA-Taiji network, respectively. The  observation periods are all set as $4$ years. The filled and open symbols represent the sBBHs that can merge within the $4$-year observation period and those cannot, respectively. The width of each symbol indicates its relative SNR ($\propto R^\gamma$ with $\gamma=1.3$), with the smallest one representing an SNR of $8$. The mock sBBHs shown here are from the sample shown by the solid histogram in Fig.~\ref{fig:SNR_LT10}. Dotted lines mark those sources with different $\mc$ and $f_{\rm i}$ but the same lifetime, e.g., $4$\,yr, $20$\,yr, and $100$\,yr, respectively.
}
\label{fig:f-M_scatter}
\end{figure}

\begin{figure}
\centering
\includegraphics[width=0.5\textwidth]{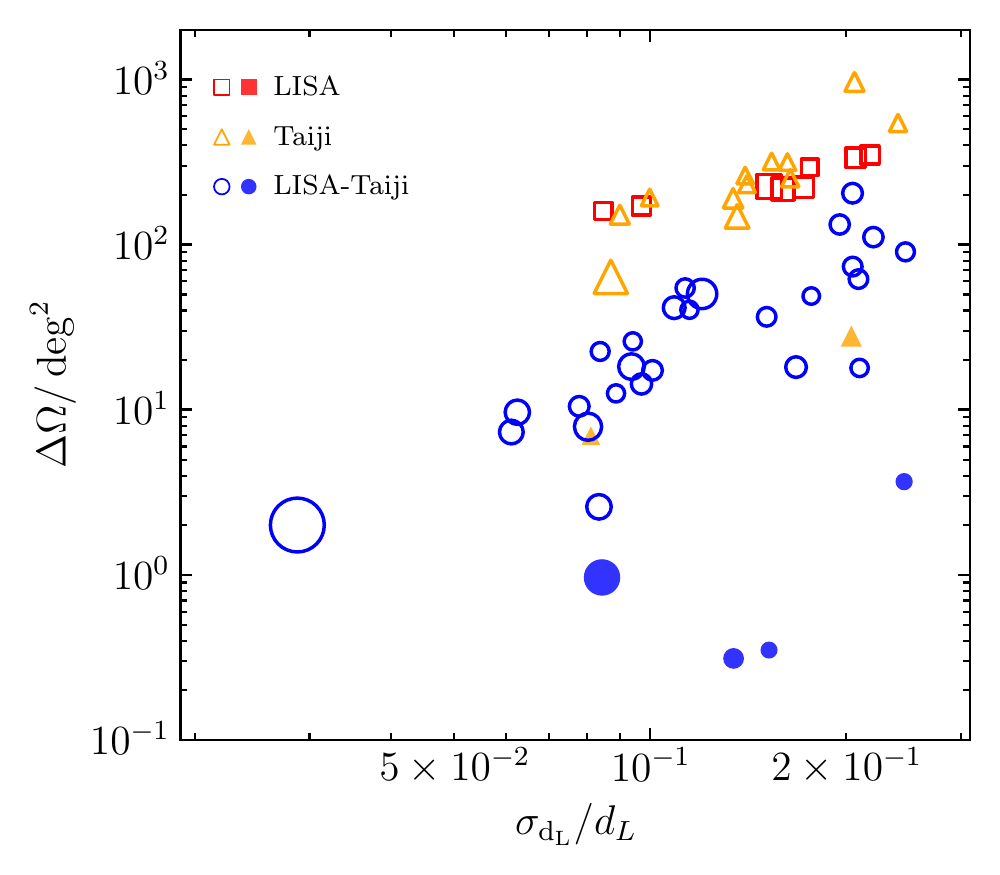}
\caption{Distribution of mock sBBHs with SNR$>8$ on the relative error of luminosity distance measurements versus the error of sky localization plane. Red squares, orange triangles, and blue circles represent those sources ``detected'' by LISA, Taiji, and LISA-Taiji network, respectively. The  observation periods are all set as $4$ years. The filled and open symbols represent the sBBHs that can merge within the $4$-year observation period and those cannot, respectively. The width of each symbol indicates its relative SNR, with the smallest one representing an SNR of $8$. The sBBHs shown here are from the three samples shown by the solid histograms in Figs.~\ref{fig:SNR_L}, \ref{fig:SNR_T}, and \ref{fig:SNR_LT}, respectively. 
}
\label{fig:err_scatter}
\end{figure}

\begin{figure}
\centering
 \includegraphics[width=0.5\textwidth]{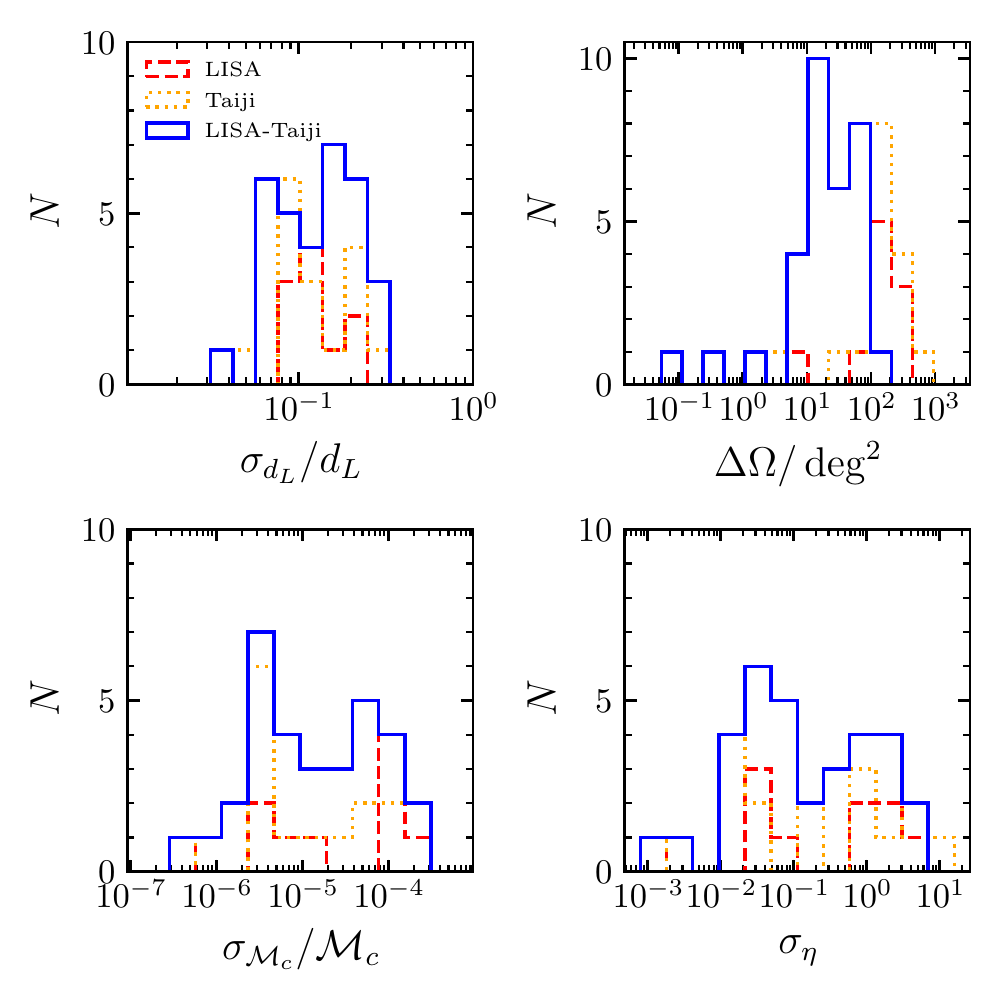}
\caption{Distribution of parameter estimation precision for mock sBBHs detected by LISA (red dashed), Taiji (orange dotted) and LISA-Taiji network (blue solid) with SNR$>8$ within a 4-year observation period. Top left, top right, bottom left, and bottom right panels show the distributions for the relative error of luminosity distance, sky localization, relative error of chirp mass, and symmetric mass ratio, respectively.
The result shown here are obtained for the three realization shown by the solid histograms in Figs.~\ref{fig:SNR_L}, \ref{fig:SNR_T}, and \ref{fig:SNR_LT}, respectively.
}
\label{fig:err_hist}
\end{figure}

Figure~\ref{fig:f-M_scatter} shows the distribution of those ``detectable'' sBBHs (defined by SNR $\varrho>8$) obtained by LISA, Taiji, and the LISA-Taiji network for the $50$-th realization on the plane of chirp mass and GW frequency at the starting time of the observations. Apparently, most of the sources (open symbols) are always in the ``inspiralling'' stage and do not move out of the sensitive frequency range of LISA and Taiji due to orbital evolution during the $4$-year observation period. We categorize these sBBHs as ``inspiralling'' sBBHs though some of them may eventually merge several tens of years later. For this type of sBBH systems, the expected SNR increases if extending the observation period. A small fraction of the ``detected'' sBBHs can merge within the $4$-year observation period (filled symbols). For the realization shown in figure~\ref{fig:f-M_scatter}, Taiji and the LISA-Taiji network can detect $2$ and $4$ sBBHs that actually merged within the $4$-year observation period. We categorize these sBBHs as ``merging'' sBBHs. Apparently, the ``merging'' sBBHs are the best sources for multi-band GW observations. 

Figure~\ref{fig:err_scatter} shows the uncertainties in the luminosity distance measurements and sky localization for those sBBHs ``detected'' by the LISA/Taiji/LISA-Taiji network with SNR $\varrho>8$ obtained from the $50$-th realization (the solid histogram shown in Figs.~\ref{fig:SNR_L}, \ref{fig:SNR_T}, and \ref{fig:SNR_LT}). As seen from this figure, the typical range for the relative errors in the luminosity distance measurements is from $\sim 0.04$ to $0.2$, and errors in localization is typically of $\sim 1\,{\rm deg}^2$ to $100\,{\rm deg}^2$. Normally, the larger the SNR, the smaller the relative errors in luminosity distance measurements and localization. The LISA-Taiji network gives much better estimations about the luminosity distances and sky localization than single LISA or Taiji. The ``merging'' sBBHs have better localization than those ``inspiralling'' sBBHs, though the SNR of the former ones are not necessarily larger than those of the latter ones.  This dichotomy is more evident if extending the observation period, say $10$ years (see Fig.~\ref{fig:err_scatter_LT10}), and we will give the reason for  this dichotomy later. 
We also show the predicted distributions for the precision for the measurements of luminosity distance (top left panel), sky localization (top right panel), chirp mass (bottom left), and symmetric mass ratio (bottom right panel) in figure~\ref{fig:err_hist}.

\begin{figure}
\centering
\includegraphics[width=0.5\textwidth]{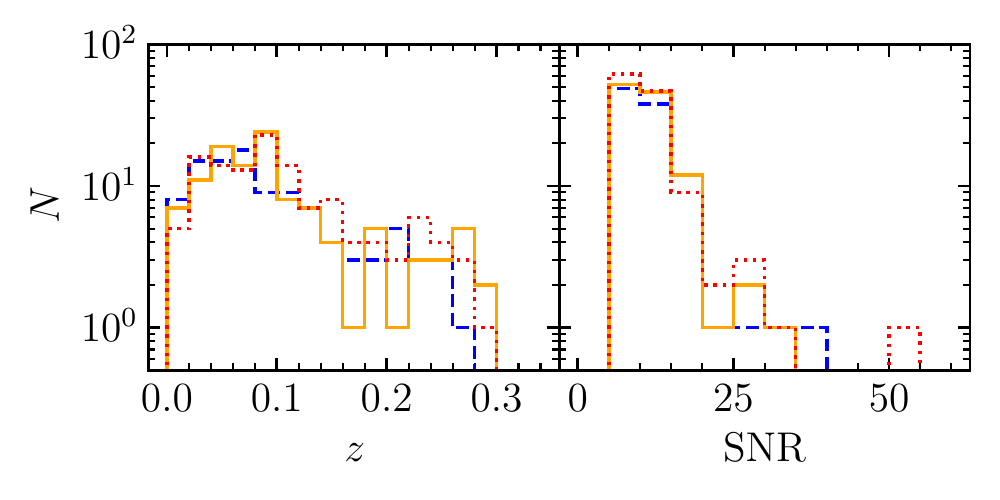}
\caption{
Redshift (left panel) and SNR (right panel) distributions of mock sBBHs detected by the LISA-Taiji network with SNR$>8$ over a $10$-year observation period. Orange solid, blue dashed, and red dotted histograms indicate the distributions of three different samples selected the same way as those shown in Fig.~\ref{fig:SNR_L}.
}
\label{fig:SNR_LT10}
\end{figure}

\begin{figure}
\centering
\includegraphics[width=0.5\textwidth]{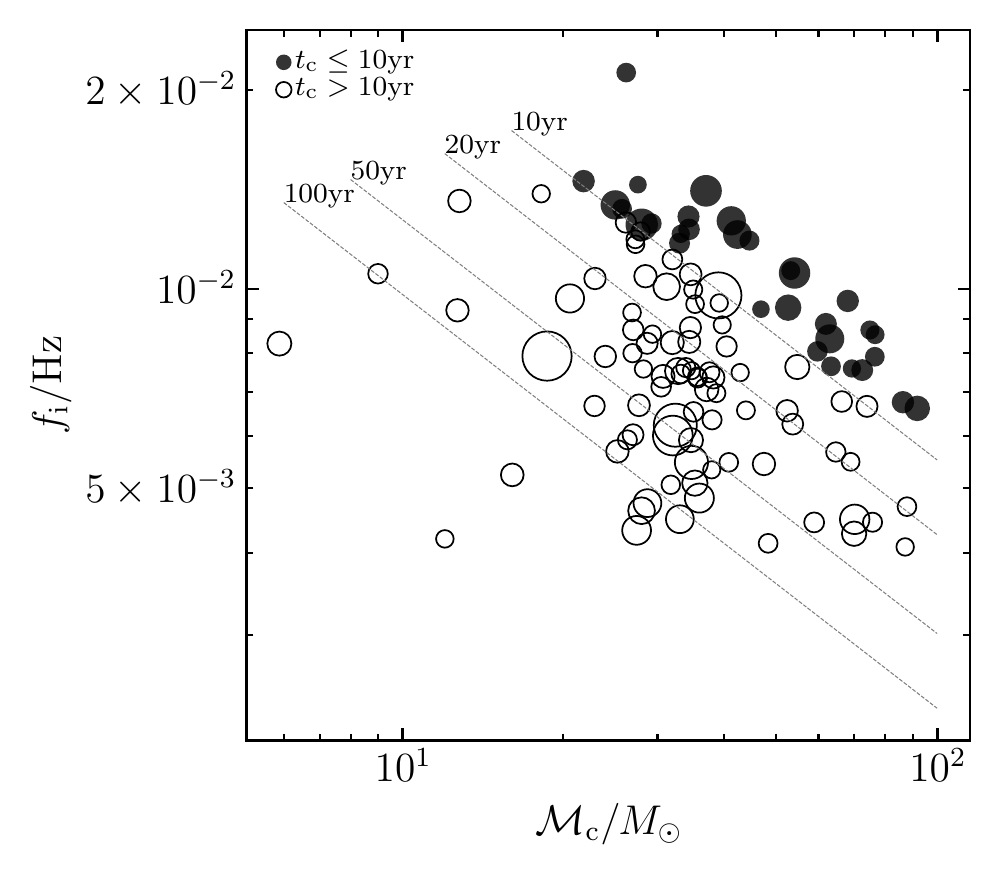}
\caption{
Distribution of mock sBBHs, ``detected'' by LISA-Taiji network with SNR$>8$ over an observation period of $10$ years, on the plane of (redshifted) chirp mass $\mathcal{M}_{\rm c}$ versus observed GW frequency at the starting time of the observation. The filled and open symbols represent the sBBHs that can merge within the $10$-year observation period and those cannot, respectively. The width of each symbol indicates its relative SNR, with the smallest one representing an SNR of $8$. The mock sBBHs shown here are from the sample shown by the solid histogram in Fig.~\ref{fig:SNR_LT10}. Dotted lines mark those sources with different $\mc$ and $f_{\rm i}$ but the same lifetime, e.g., $10$\,yr, $20$\,yr, $50$\,yr, and $100$\,yr, respectively.
}
\label{fig:f-M_scatter_LT10}
\end{figure}

\begin{figure}
\centering
\includegraphics[width=0.5\textwidth]{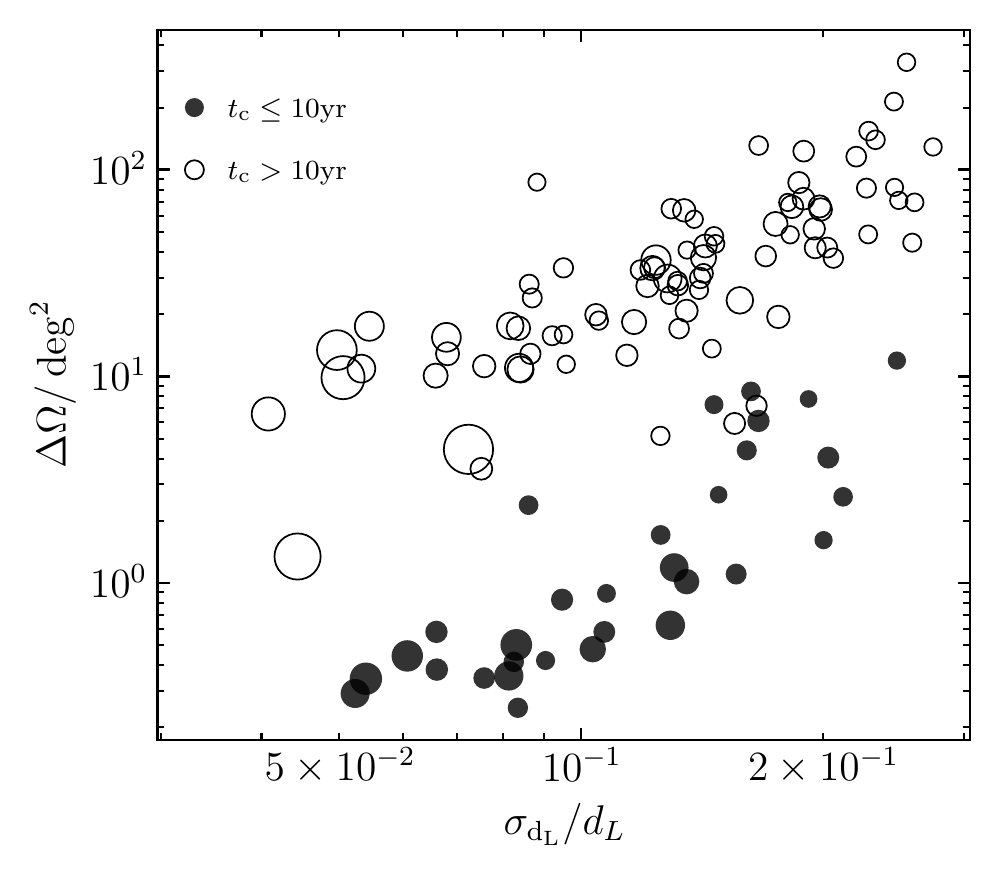}
\caption{
Distribution of mock sBBHs, ``detected'' by the LISA-Taiji network with SNR$>8$ over an observation period of $10$ years, on the relative error of luminosity distance measurements versus the error of sky localization plane.  The filled and open symbols represent the sBBHs that can merge within the $10$-year observation period and those cannot, respectively. The width of each symbol indicates its relative SNR, with the smallest one representing an SNR of $8$. The mock sBBHs shown here are from the sample shown by the solid histogram in Fig.~\ref{fig:SNR_LT10}.
}
\label{fig:err_scatter_LT10}
\end{figure}

\begin{figure}
\centering
\includegraphics[width=0.5\textwidth]{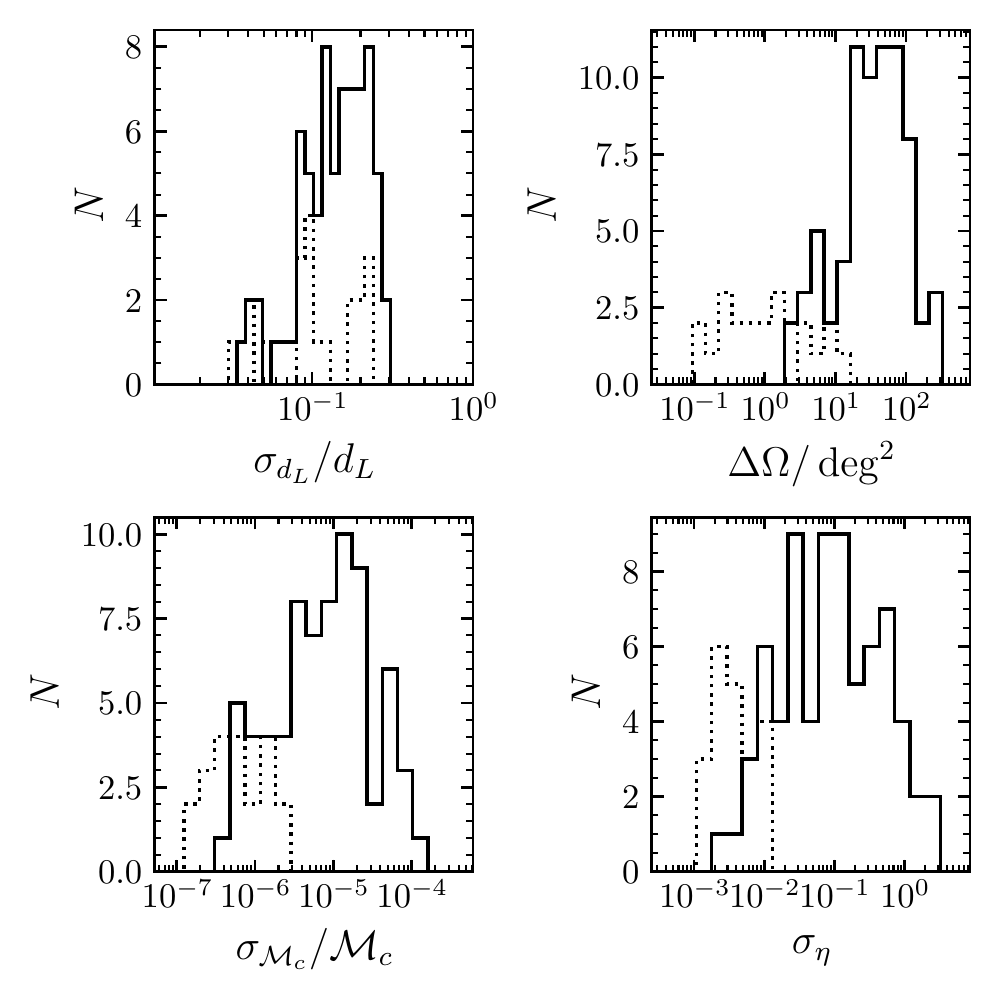}

\caption{
Distribution of parameter estimation precision for mock sBBHs detected by LISA-Taiji network with SNR$>8$ over 10 years observation. Top left is relative error of luminosity distance, top right is sky localization, bottom left is relative error of chirp mass and bottom right is symmetric mass ratio. The dotted and solid lines represent the sBBHs that can merge within the 10-year observation period and those cannot, respectively.
The results shown here are from the three realizations shown by the solid histograms in \ref{fig:SNR_LT10}.
}
\label{fig:err_hist_LT10}
\end{figure}

If the mission lifetime extend to 10 years, the result would be more optimistic.
Assuming a $10$-year observation of the LISA-Taiji network, the number of sBBHs that have SNR$>8$ is significantly increased, as shown in Figure \ref{fig:num}. The expected number of detection is $N(\varrho>8) = 114_{-10}^{+11}$.
Figure~\ref{fig:SNR_LT10} shows the redshift and SNR distributions of these sBBHs for three realizations selected from the total $100$ realizations ranking by the total number of sBBHs with SNR$>8$, and they are the $16$-th, $50$-th, and $84$-th realizations, respectively. As seen from the figure, the maximum redshift of these sBBHs is roughly $z\sim 0.3$. If adopting SNR $\varrho>15$ as a conservative ``detection'' threshold for sBBHs by the LISA-Taiji network, then we still expect to detect a significant number of sBBHs, i.e., $17^{+6}_{-4}$ (see Fig.~\ref{fig:num}).

Figure~\ref{fig:f-M_scatter_LT10} shows the distribution of those sBBHs ``detected'' by the LISA-Taiji network with SNR$>8$ obtained from the $50$-th realization (the solid histogram show in Fig.~\ref{fig:SNR_LT10}). As seen from this figure, a fraction $\sim 1/4$ of those sBBHs ($31$ out of $114$) can merge within the $10$-year observation period, and they all have large chirp mass $\gtrsim 20M_\odot$. While the majority of those sBBHs cannot merge within the $10$-year observation period, which is consistent with \citet{Sesana2017}. These ``inspiralling'' sBBHs have relatively smaller frequencies at the starting time of the observation, and a few of them can have quite small chirp mass $<10M_\odot$. The ``inspiralling'' ones can have large SNR because: (1) they are ``monitored'' in the whole observation period; (2) their frequencies at the starting time of observations are in the most sensitive range of the detectors; and (3) they are relatively closer than those sBBHs that can merge within the observation period. If set $\varrho_{\rm th}=15$ as the threshold, then the expected detection numbers of ``merging'' and ``inspiralling'' sBBHs are $3_{-1}^{+3}$ and $13_{-3}^{+5}$, respectively.

Figure~\ref{fig:err_scatter_LT10} shows the uncertainties in the luminosity distance measurements and sky localization for those sBBHs ``detected'' by the LISA-Taiji network with SNR$>8$ obtained from the $50$-th realization (the solid histogram show in Fig.~\ref{fig:SNR_LT10}). As seen from this figure, the ``merging'' sBBHs can have better sky localization, in the range from $\sim 0.3\,{\rm deg}^{2}$ to $10\,{\rm deg}^2$, while ``inspiralling'' sBBHs have poorer sky localization, typically $\gtrsim 10\,{\rm deg}^2$. The main reason is that the mass determination for ``merging'' sBBHs is more accurate than the ``inspiralling'' sBBHs because of more significant changes of GW frequencies for ``merging'' sBBHs relative to the ``inspiralling'' ones, as shown in figure \ref{fig:err_hist_LT10}. The relative errors in the luminosity distance measurements of ``merging'' and ``inspiralling'' sBBHs are similar and typically in the range from $\sim 0.04$ to $\sim 0.2$. The predicted distributions for parameter estimation precision are also shown in figure~\ref{fig:err_hist_LT10}.

\section{Conclusions and Discussions}
\label{sec:conclusions}

In this paper, we investigate the possibility of ``detect'' sBBHs by future space-based interferometer GW detectors, such as LISA, Taiji, and especially the network formed by LISA and Taiji. By adopting the sBBH merger rate density, its evolution, and its dependence on the sBBH properties constrained by LIGO/VIRGO observations, we randomly generate mock samples for sBBHs at low redshift and estimate the SNR for these mock sBBHs and the uncertainties in the parameter determinations from GW signals. Assuming an observation period of $4$ years, we find that LISA and Taiji may detect $8^{+3}_{-2}$ and $14^{+4}_{-3}$ with SNR\,$>8$ (or $1^{+1}_{-1}$ and $2^{+2}_{-1}$ with SNR\,$>15$), respectively. Among them, $0^{+1}_{-0}$ (or $0^{+0}_{-0}$)
for LISA and $2^{+1}_{-1}$ (or $0^{+1}_{-0}$) for Taiji can merge within the observation period. If combining LISA and Taiji observations as a network, then the number of sBBHs with SNR\,$>8$ (or $>15$) is $30^{+7}_{-5}$ (or $5^{+6}_{-2}$), and among them $3^{+2}_{-1}$ (or $0^{+1}_{-0}$) can merge during the observation period. If extending the observation period to $10$ years, then the LISA-Taiji network may detect $114^{+11}_{-10}$ (or $17^{+6}_{-4}$) sBBHs with SNR\,$>8$ (or $>15$), among them only $23^{+6}_{-3}$ (or $3^{+3}_{-1}$) can merge within the observation period. Our results suggest that the LISA-Taiji network may be able to detect at least a handful to more than twenty sBBHs even if assuming a conservative SNR threshold ($>15$) for ``detection'', which means that the LISA-Taiji network will grant the multi-band GW observations of sBBHs. However, the detection of individual sBBHs by LISA or Taiji only seems not so optimistic within an observation period of several years. 

Note here that we consider the perspective of sBBH detection by space detector only in this paper. However, one may also search for sub-threshold events, with known parameters determined by ground-based GW observatories, in the archived data of space observation. With this strategy, the required SNR threshold for space detection may be lowed to $\sim4$, as pointed out by \citet{Wong2018} and \citet{Ewing2021}. Consequently, the number of GW sources for multiband studies may increase significantly, e.g., by a factor of $4$-$8$. This suggests that the multiband GW astronomy is quite promising, which enables strong constraints on gravity theories and/or the formation channels of sBBHs \citep[e.g.,][]{Sesana2017, Cutler2019}.

We also further estimate the uncertainties in the estimations of these sBBH system parameters from GW signals. We find that the relative errors in the luminosity distance measurements and sky localization are mostly in the range of $0.05-0.2$ and $1-100\deg^2$, respectively, for sBBHs detected by the LISA-Taiji network. Among the ``detected'' sBBHs, those that can merge within the observation period have relatively better mass measurement and localization, and the localization is typically smaller than several ${\rm deg}^2$.
%

We also note here that all the numbers obtained in the present paper are based on the local merger rate density, its evolution, and its dependence on sBBH parameters. However, the current constraints obtained from the LIGO/VIRGO O3a observations still have some uncertainties. For example, the local merger rate density is $19.1^{+16}_{-10}{\rm Gpc^{-3}\,yr^{-1}}$, its uncertainty is close to a factor of $2$. Such an uncertainty would introduce more or less a factor of $2$ to the estimated numbers for sBBHs ``detection'' by LISA, Taiji, and the LISA-Taiji network.

\normalem
\begin{acknowledgements}
This work is partly supported by the National Key R\&D Program of China (Grant Nos. 2020YFC2201400, 2020SKA0120102, and 2016YFA0400704), the National Natural Science Foundation of China (Grant Nos. 11690024, 11873056, and 11991052), the Strategic Priority Program of the Chinese Academy of Sciences (Grant No. XDB 23040100), and the Beijing Municipal Natural Science Foundation (Grant No. 1204038).
\end{acknowledgements}
  
\bibliographystyle{raa}

\end{document}